\documentclass[aps,pre,superscriptaddress,showpacs,floatfix]{revtex4}

\usepackage{epsfig}
\usepackage{amsfonts}
\usepackage{amssymb}
\usepackage{newlfont}

\DeclareMathAlphabet{\mathitbf}{T1}{cmr}{bx}{it}

\begin{document}

\title{Numerical Simulations of Random Phase Sine-Gordon Model
and Renormalisation Group Predictions}

\author{D. J. Lancaster} \affiliation{Department of Computer Science, 
Westminster University, London, UK}  
\email{ D.J.Lancaster@westminster.ac.uk}

\author{J. J. Ruiz-Lorenzo} \affiliation{Depto. de F\'{\i}sica,
Facultad de Ciencias, Universidad de Extremadura, 06071 Badajoz,
Spain.\\ Instituto de Biocomputaci\'on y F\'{\i}sica de los Sistemas
Complejos (BIFI).}  \email{ruiz@unex.es}

\date{\today}

\begin{abstract}
  Numerical Simulations of the random phase sine-Gordon 
  model suffer from strong finite size effects preventing the
  non-Gaussian $\log^2 r$ component of the spatial correlator 
  from following the universal infinite volume prediction. 
  We show that a finite size prediction based on perturbative
  Renormalisation Group (RG) arguments agrees 
  well with new high precision simulations for small coupling and 
  close to the critical temperature.
\end{abstract}


\maketitle

\section{Introduction}

The random phase sine-Gordon model plays a role in understanding a
variety of phenomena.  The model has been physically interpreted in
terms of roughening of crystalline surfaces \cite{toner-divicenzo},
pinning of flux lines in superconducting film
\cite{Fisher,Hwa2,tsai-shapir,fisher-hwa,Hwanew,Bolle} charge density waves
\cite{Fukuyama-Lee,narayan-fisher}, and other systems.  The model has
generated a variety of theoretical predictions
\cite{fisher-hwa,GLD1,toner-divicenzo}, not all in agreement, and
continues to be a scene of activity
\cite{ARSANBI,SANBIMO,SLD,Rieger3}.  Numerical simulations
\cite{batrumi-hwa,cule-shapir,Rieger1,Rieger2,US1,US2,Zeng,SANBIMO,Rieger3}
might be expected to play a decisive role in this situation, but this
has not in fact been the case.  Over the past few years, some of the
differences between simulations and theory have been resolved as a
result of better simulations and analysis \cite{cule-shapir,US1} and
corrections to the precise values of theoretical predictions
\cite{CLD,nattermann-scheidl}, but there still remain discrepancies.

\begin{figure}
\begin{center}
\leavevmode
\epsfig{file=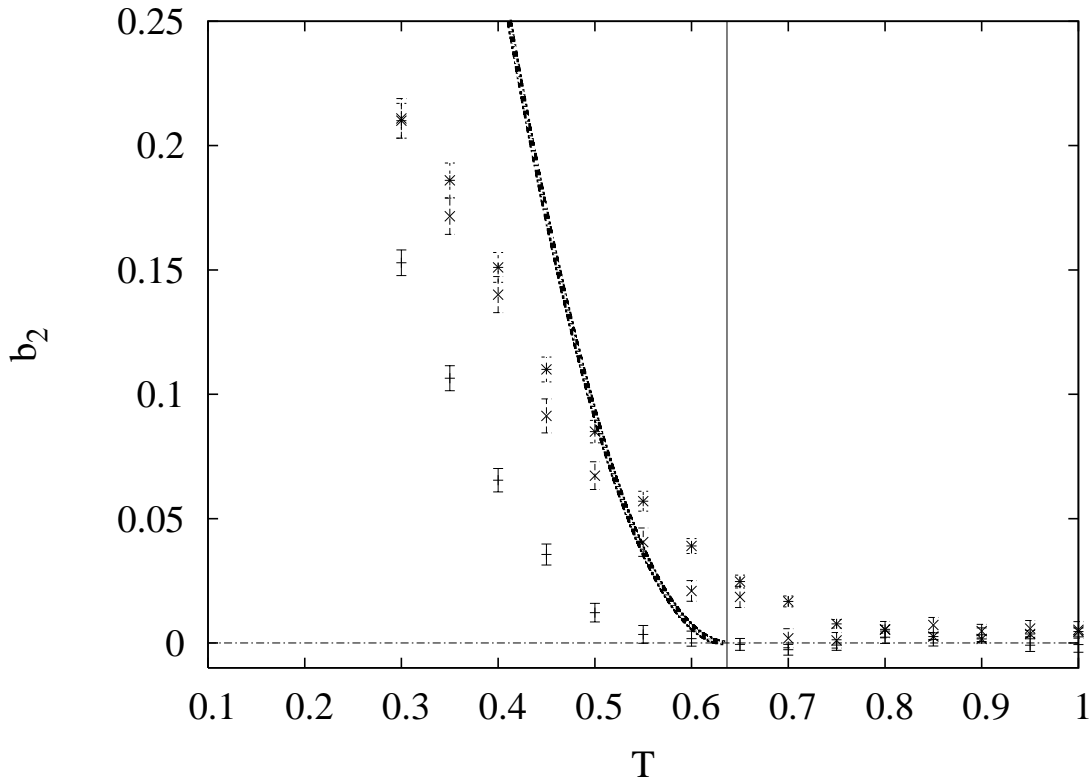,width=0.6\linewidth,angle=270}
\end{center}
\caption{
Numerical results for the coefficient $b_2$ (see the text) of the 
$\log^2 r$ term in the correlator on a $64^2$ lattice
using data originally discussed in \cite{US1,US2}. 
Bottom to top:
$\lambda=0.5, 2.0$ and the discrete height model (formally equivalent to the
sine Gordon model with $\lambda=\infty$).  
The vertical line indicates
the position of the infinite volume critical temperature
at $T_c=2/\pi$, and the horizontal line shows $b_2=0$. The wider  line
marks the (universal) infinite volume RG prediction ($b_2=2 \tau^2$).
}
\label{fig:old}
\end{figure}

The primary discrepancy between theory and simulation concerns the presence of
a $\log^2 r$ component of the spatial propagator computed at distance $r$ in
the {\it high} temperature phase (since the high temperature phase is
Gaussian, the spatial correlator should be  simply $\log r$).  
In the renormalisation group (RG) approach,
this component is expected below the critical temperature at length scales
beyond a characteristic length that diverges as the critical temperature is
approached from below.  However, in finite lattice simulations, 
this component is observed
\cite{US2} above the expected critical temperature, depending on the strength
of the disorder, and leads to uncertainty in actually identifying the location
of the critical point~\cite{SANBIMO}.  Although the possibility that the
critical temperature could depend on the strength of the disorder was raised
in theoretical work~\cite{kierfeld}, the accepted theoretical view is that
this does not occur.  The reason for this is a symmetry of the Hamiltonian,
noted by Cardy and Ostlund~\cite{cardy-ostlund} and recognised as a particular
example of the so-called statistical tilt symmetry \cite{SCVIBEOR}. 
Another discrepancy concerns the strength
of the $\log^2 r$ component in the {\it low} temperature phase.
The RG prediction for this strength turns out to be universal, yet
it does not match the value observed in simulations on achievable
lattice sizes. Figure (\ref{fig:old}) illustrates the discrepancies
between the infinite volume RG prediction and data from simulations
of several years ago~\cite{US2}.
These phenomena have been observed in other numerical studies~\cite{SANBIMO}.

In this paper we use perturbative RG arguments to estimate the size of the 
$\log^2 r$ component for small couplings in the vicinity of the
critical temperature  bearing in
mind the finite size of the lattices used in simulation.
We confront these estimates with data from
new high precision simulations made within the restricted 
parameter range where perturbation theory can be relied upon.
Our estimates {\it do} display a
$\log^2r$ behaviour above the critical temperature and match the 
strength below the critical temperature therefore
resolving the discrepancy.

To use the RG theory to make predictions on the lattice, 
we need to compute
the coefficients appearing in the RG equations using a lattice regulator.  The
usual treatment is via a Coulomb gas \cite{cardy-ostlund} which involves a
completely different type of regulator.  We have obtained RG coefficients for
the lattice regulator by adapting Kogut's original approach of momentum shell
integration~\cite{Kogut}. These coefficients are substantially smaller than
the conventional Coulomb gas ones, and allow  quantitative
agreement with the simulation data.  Nonetheless, the RG theory is only valid
for small couplings, and only in this region do we obtain good agreement.

We commence with a description of the model and simulation data we wish 
to explain but relegate technical details of numerical work to appendix A.
Then in section III, we briefly review the RG for this model and refer
to appendix B for the full computation of RG coefficients appropriate
to the lattice regulator.
Section IV discusses our technique for estimating the finite size correlator 
and presents results in the form of a figure confronting the 
prediction against the new simulation data.
The conclusion critically discusses our treatment and since this 
work does not constitute a full finite size theory, it also discusses 
some of the issues involved in such a theory.  

\section{Simulations}

The random phase sine-Gordon model has Hamiltonian,
\begin{equation}
H = \sum \,
(\phi_i - \phi_j)^2
-\lambda \sum \cos 2\pi(\phi_i - \eta_i)\,.
\label{Hamilt}
\end{equation}
Depending on the physical system, the
continuous field $\phi$  may be interpreted in different ways.
In the case of a surface, $\phi$ is the height of the interface so
the field $\phi$ is not periodic and vortices are excluded.
The disorder is contained in the term $\eta$ 
which is characterised by some probability distribution.
Usually a flat distribution for the quenched disorder in $0\le \eta < 1$, 
is assumed,
but other distributions have also been simulated~\cite{SANBIMO}.

In order to clarify the problems that arise in simulations,
let us start by recalling the results of our earlier 
numerical study \cite{US2} 
of this model which form the basis of figure (\ref{fig:old}).
Signals of a transition were clear in 
both the static and dynamic properties
\footnote{In numerical work, correlation functions are the basis
for these signals, but Fisher and Hwa \cite{fisher-hwa} pointed out that they
are inappropriate for experiment since long range correlations of the
disorder can confuse their interpretation and therefore they
proposed the sample dependent susceptibility. Experimentally the
curves $B(H)$ have been
measured \cite{Bolle}.}.
Although the static correlation function 
had the expected $\log r$ behaviour for high temperature, it was
clear that another component was present at low
temperatures. The results were not sufficiently sensitive
to determine the functional form of this additional
component, but a $\log^2 r$ term as predicted by the
RG studies allowed a good fit to the data. 
To be precise, the simulated spatial correlator was fitted with the function
(using $\langle (\cdot \cdot \cdot) \rangle$ to denote average over the thermal noise and
$\overline{(\cdot \cdot \cdot)}$ as the average over the quenched disorder):
\begin{equation}
G(r) \equiv \overline{\langle \left(\phi_r - \phi_0  \right)^2  \rangle}=  b_1 P_L(r)
+ b_2 P_L^2(r)
\label{fitform}
\end{equation}
$b_1$ and $b_2$ are the fit parameters and $P_L(r)$ is
the Gaussian correlator on a lattice of size $L$,
\begin{equation}
  P_L(r)  = \frac{1}{2 L^2} \sum_{n_1=1}^{L-1} \sum_{n_2=0}^{L-1}
  \frac{1-\cos(\frac{2 \pi r n_1}{L})}
  {2-\cos(\frac{2 \pi n_1}{L})-\cos(\frac{2 \pi n_2}{L})}
   \simeq
   {1\over {2\pi}} \log({2 \sqrt{2} e^{\gamma_E}}r)\,,
\label{eq:PsubL}
\end{equation}
and  the symbol $\simeq$ holds for large lattices, 
$L\gg 1$ and
length scales, $r$, that are large but not approaching $L$.
$\gamma_E$ is the Euler-Mascheroni constant.
This is actually the correlator along a lattice axis
and is the appropriate quantity to compare with the correlators
obtained from simulation.

Instead of relying on our old data to make comparison with our predictions
we have performed new high precision simulations.
These new simulations at small coupling, $\lambda=0.5$, 
remain on the same $64^2$ lattice but
have better thermalization, an improved random number generator
and higher statistics. Details of the simulations and comparison
with the old data are discussed in Appendix A along with
additional data exploring other values of $\lambda$.
Data for the coefficient $b_2$ of the $\log^2 r$ term based on 
new simulations are presented in figure (\ref{fig:new}) later in the paper.

\section{Renormalisation Group}

A convenient basis for the RG studies is the continuum 
replicated version of the
Hamiltonian (\ref{Hamilt}) following Cardy and Ostlund~\cite{cardy-ostlund}:
\begin{equation}
\beta H = \int d^2x \,{1\over 2}
\sum_{\alpha\beta}K_{\alpha\beta} \partial \phi_\alpha\partial \phi_\beta
-{g\over a^2} \sum_{\alpha\beta} \cos 2\pi(\phi_\alpha - \phi_\beta)\,.
\label{eq:repH}
\end{equation}
Greek letters denote the $n$ replica indices and we have 
made $g$ dimensionless by explicitly
writing the UV cutoff $a$.

The kinetic term of the replicated Hamiltonian is
parameterised with separate coefficients for the on and off-diagonal pieces,
$K_{\alpha\beta} =   K\delta_{\alpha\beta} 
+ (K-\tilde{K})(1-\delta_{\alpha\beta})$,
since the off-diagonal piece appears after renormalisation even if not present
initially.
The bare Hamiltonian only has a diagonal term, so the bare parameters are:
$\tilde{K}(0) = K(0) = 2\beta$.
The bare value of $g$ is related to the parameters of the 
simulated Hamiltonian by: $g(0) = (\beta\lambda/2)^2$.

RG equations may be obtained from  (\ref{eq:repH}) 
\cite{cardy-ostlund,goldschmidt-houghton,Goldschmidt-Schaub,DENNIS,CLD,ERG},
though the same results are obtained without replicas \cite{fisher-hwa}. 
In either
case an additional term is generated: in the replica calculation
it appears as the off-diagonal part of $K_{\alpha\beta}$, whereas
in the other approach it appears as the coefficient of
a new term in the Hamiltonian, linear in the derivative of $\phi$.
These same equations have been derived in a variety of ways;
a Coulomb gas approach \cite{cardy-ostlund}, 
conventional field theory \cite{goldschmidt-houghton}, 
conformal field theory \cite{DENNIS},
with the
help of the exact renormalisation group \cite{ERG}
and in appendix A, we ourselves use momentum shell integration.
Irrespective of the method of derivation, the parameters
change with the RG scale, $l$, in the following way:
\begin{eqnarray}
\frac{ d g}{dl} &=& 2\tau g - C g^2  \,,\\
\frac{ d K}{dl} &=& -A g^2 \,,\\
\frac{ d \tilde{K}}{dl} &=& 0 \,.
\end{eqnarray}
Where the reduced temperature 
$\tau = (1-\tilde{K}_c/\tilde{K}) = (1 - T/T_c)$.
The critical temperature is at $T_c = 2/\pi$ (hence, ${\tilde K}_c=\pi$).
These perturbative RG equations have been obtained in a double expansion: 
small coupling $g$ and 
small reduced temperature $\tau$ \cite{goldschmidt-houghton}.
They are valid for small couplings in the region near the critical 
point where higher order terms can be ignored.

The asymptotic form as $l \to \infty$ of
the solution of these equations depends on the phase.
In the high temperature region where $\tau$ is negative,
the fixed point value of
$g(l)$ is zero and $K(l)$ goes to a constant.
In the low temperature region 
$g(l)$ becomes $g^* = 2\tau/C$, 
and the coupling $K(l)$ 
behaves as $K(l) = -4(A/C^2)\tau^2 l$.
The consequences of these solutions
are fully explored in the literature and we do not
repeat the analysis here.

The RG coefficients $A, C$ are dimensionless, non-universal
constants of order unity. 
They depend on the temperature but 
various workers~\cite{fisher-hwa,goldschmidt-houghton} 
noted that the amplitude ratio $A/(\tilde{K}C)^2$ appearing in the
asymptotic form of $K(l)$ below $T_c$,
is universal at the critical point.
The correct value for this ratio was clarified by Carpentier and Le
Doussal \cite{CLD} (see also reference~\cite{nattermann-scheidl}).  In
our conventions (note that we keep a factor of $2\pi$ inside the
cosine term), this ratio is:
\begin{equation}
{A\over (\tilde{K}C)^2} = 
{R \over(2\pi)^2}{T\over T_c} =
{1 \over(4\pi)}{T\over T_c}\,.
\end{equation}
$R$ is the universal ratio defined in \cite{CLD}
and carefully shown to take the value $\pi$.

The non-universal values of $A$ and $C$ themselves are necessary to compare
with simulation. 
The regulator natural in the Coulomb gas approach used by
Cardy and Ostlund~\cite{cardy-ostlund} gives values 
$ C = 4\pi$ and $A = 4\pi^3$. These values are
rather high and we should properly use values from a lattice
regulator. A full perturbative RG computation on the lattice becomes difficult
so we have resorted to the following argument.  

We have used  momentum shell integration 
following Kogut \cite{Kogut} to rederive the RG equations.
This technique needs a modification first pointed out in 
\cite{KnopsDenOuten} to take account of the
correct operator product expansion, however all computations
can be consistently expressed in terms of Gaussian integrals without
any need to be concerned with the subtleties of two dimensional
massless canonical field theory. 
Details are given in appendix B, and this technique yields
an expression in terms of the Gaussian propagator 
$<\phi_\alpha(r)\phi_\beta(0)>=K^{-1}_{\alpha\beta}G_0(r)$:
\begin{equation}
C = 
{8\pi^2\over {\tilde K}}
\Lambda^2\int d^2\xi \, 
\Lambda {dG_0(\xi)\over d\Lambda}
e^{{4\pi^2\over{\tilde K}} \left[ G_0(\xi) - G_0(0)\right]}
=
4\pi e^{\Phi(\infty)}\,.
\end{equation}
The final expression on the right is evaluated at the critical point,
is identical that obtained by a different 
procedure in \cite{CLD} and
gives the correct value of the universal ratio.
$\Phi$ depends on the regulation scheme and 
is defined in terms of the subtracted propagator:
\begin{equation}
\left[ G_0(r) - G_0(0)\right] 
= - {1\over 2\pi} \log (r\Lambda) + {1\over 4\pi} \Phi(r\Lambda) \,. 
\end{equation}
The momentum shell integration technique requires a regulator
that is sufficiently smooth to make the term ${dG_0(\xi)\over d\Lambda}$
short range and thus render the expression for $C$  well defined.
The lattice propagator has this property,
and we can identify $\Phi(\infty)$ using the asymptotic
form of the lattice regulator (\ref{eq:PsubL})~\cite{ID}:

\begin{equation}
\left[ G_0(r) - G_0(0)\right]_\mathrm{lattice} 
\rightarrow - {1\over 2\pi} \log (r\Lambda 2\sqrt{2}e^{\gamma_E})\,.
\end{equation}
We therefore use the value of $C$ given by
\begin{equation}
C = {\pi\over 2}e^{-2\gamma_E}\,.
\end{equation}
Recalling the value of the Euler-Mascheroni constant
$\gamma_E\simeq 0.577...$, we find 
this value of $C$ to be substantially smaller than the
Coulomb gas value.

\section{Finite Size}

In order to compare the RG theory with simulation, we need
the correlation function on a finite geometry. 
The usual derivation of the RG correlator in infinite volume was given 
by Toner and Vicenzo~\cite{toner-divicenzo} and
we shall proceed in a similar way {\it without} taking
asymptotic values for the RG scale. 
Numerically it is better to work in the real space, 
but the fields $\phi$ do not transform simply under the RG, so we
consider the composite fields $e^{iq\phi}$ that have better 
properties~\cite{goldschmidt-houghton}.
General RG arguments allow us to write an equation for the
correlation function $C_q(r)$ of these vertex operators (see for example
reference~\cite{goldschmidt-houghton}):
\begin{equation}
C_q(r,\tilde{K},K(0),g(0))\equiv 
\langle \exp\left(i q (\phi(r)-\phi(0)\right) \rangle 
= \exp\left(-\int^1_{1/r} \gamma_q(g(x))
  \frac{dx}{x}\right) C_q(1,\tilde{K},K(\log r),g(\log r))\,.
\label{rg_eq}
\end{equation}
Where $\gamma_q(g)$ is twice the (RG) dimension
of the vertex operator $\exp(i q \phi(r)$. The leading terms in
the perturbative expansion of this quantity are:
\begin{equation}
\gamma_q=\frac{q^2}{2 \pi} {1\over \tilde K}
\left(2-\frac{K}{\tilde K} \right) +\gamma_2(q) g^2 \,.
\end{equation}
Determining the value of $\gamma_2$, twice the anomalous dimension, in our RG
scheme is not straightforward (it requires the OPE of three vertex operators)
and neither has it been computed in other renormalisation schemes.  We have
therefore only kept the leading term in $\gamma_q$ when expanding
equation (\ref{rg_eq}) in $q$.
We expand to order $q^2$ and write $C_q(r)=1-\frac{q^2}{2} G(r)+O(q^4)$, to 
obtain an RG equation for $G(r)$:
\begin{equation}
G(r,\tilde{K},K(0),g(0))
= G(1,\tilde{K},K(\log r),g=0)+2 \int^{\log(r)}_0 dl \frac{1}{2 \pi \tilde K}
  \left(2-\frac{K(l)}{\tilde K} \right) \,.
\label{Eqcorrg2}
\end{equation}
This expression amounts to using the zeroth order formulae improved
with the running coupling constants computed to second order in perturbation
theory and is the same procedure usually followed.
We will show that this approximation provides us with the same
qualitative picture obtained in numerical simulations and a good
agreement with the numerical data obtained at small coupling constant. 
Besides omitting the order $g^2$ term from the renormalisation 
factor $\gamma_q$, we have consistently terminated the 
perturbative expansion of $G$ at zeroth order.
In fact we have obtained an expression for the order $g^2$ contribution
to $G$, but the resulting formula is not amenable to
numerical calculation on suitable size lattices~\footnote{The
first order term does not vanish on a finite lattice, but it
can be evaluated numerically and its effect is negligible.}.
We remark that even if we were able to include these order $g^2$ 
terms, it would only be a step towards greater precision since 
the complete model on the
lattice also requires irrelevant operators (see appendix C).

On the lattice we replace $\log r$ by its lattice version
equation (\ref{eq:PsubL}), 
$\log(r) \to {2 \pi} (P_L(r)-P_L(1))$,
so our final equation for the correlator is:
\begin{equation}
G(r,\tilde{K},K(0),g(0))
= \frac{2}{\tilde K}\left(2-\frac{K\left(2 \pi[P_L(r)-P_L(1)]\right)}{\tilde K}\right)
P_L(1) 
+2 \int^{2 \pi (P_L(r)-P_L(1))}_0 dl \frac{1}{2 \pi \tilde K}
  \left(2-\frac{K(l)}{\tilde K} \right) \,.
\label{eq:RGcorrRSF}
\end{equation}

As is the case for the momentum space version, 
this equation depends on the running coupling $K(l)$
and consequently the behaviour of the correlator  
depends on the phase. The standard, infinite volume, analysis
is based on using the asymptotic solution for $K(l)$.
In the high temperature phase where $K(\infty)$ is constant this
yields a $\log r$ correlator with  
the correct coefficient ($T/2\pi$) even at temperatures beyond
the region of validity of the perturbative result.
At low temperature the asymptotic solution 
$K(l) \to -4(A/C^2)\tau^2 l$ leads
to the $\log^2r$ behaviour~\cite{toner-divicenzo}.
\begin{equation}
G(r,\tilde{K},K(0),g(0))=
2 \tau^2 P^2_L(r) + ...
\label{eq:RGcorrasy}
\end{equation}
This leading term has universal coefficient, which takes a value $b_2 =
2\tau^2$ (see equation (\ref{fitform})).  The sub-leading $\log r$ term, 
(the dots in the equation above)
can be obtained from corrections to the
asymptotic value of $K(l)$, though its precise value is dependent on the
regulation scheme.  Comparison of the two terms yields a length scale that
separates the $\log r$ from the $\log^2r$ behaviour.  Since the coefficient of
the $\log$ does not have any $\tau$ factor, this crossover scale diverges as
$e^{\mathrm{const}/ \tau^2}$ as $T_c$ is approached from below. This is the
problem identified in the Introduction.

To make predictions for finite size, we use the full solution of the
RG equation for $K(l)$~\cite{CLD}: 
\begin{equation}
K(l) = K(0) - D\tau\left( 
\log \left( 1 + \chi(e^{2\tau l} - 1)\right)
-\chi(1-\chi){e^{2\tau l} - 1 \over
1 + \chi (e^{2\tau l} - 1)} \right) \,.
\label{eqn:Kofl}
\end{equation}
Where the parameter $\chi = g(0)C/2\tau$ has the same sign as $\tau$.  $D =
2A/C^2 = 1/T$, is universal, as discussed in the previous section.  This is an
exact solution of the perturbative RG equations valid in the region of small
$g(l)$, namely for small $\lambda$ near the transition.  By using this
expression and numerically solving the real space RG equation (i.e. computing
numerically the integral which appears in the right of equation
(\ref{eq:RGcorrRSF})), we are able to compute the finite size lattice
correlator $G(r)$.  The resulting correlator could be compared directly
against simulation data for this quantity, but it is clearer to present the
comparison in the same way as we did in our earlier work by fitting the
predicted correlator with a form (\ref{fitform}).

\begin{figure}
\begin{center}
\leavevmode
\epsfig{file=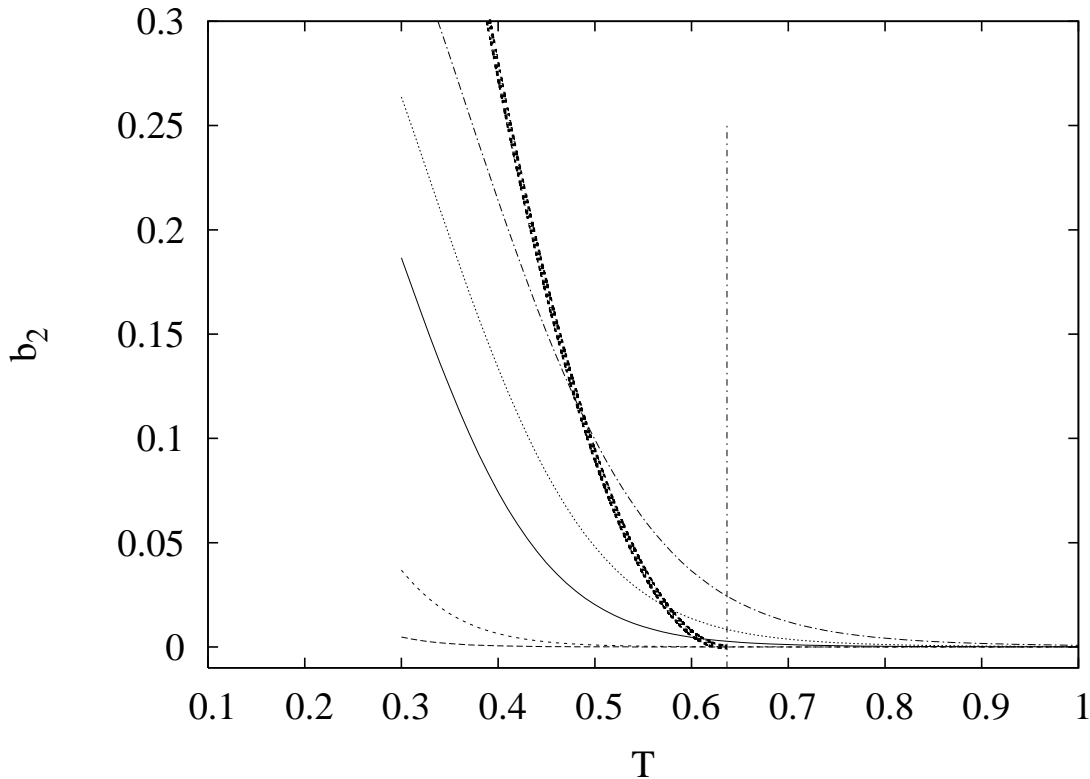,width=0.6\linewidth,angle=270}
\end{center}
\caption{
Theoretical estimate for the coefficient $b_2$ of the 
$\log^2 r$ term in the correlator 
as a function of the temperature as determined by fitting
RG predictions for a lattice size $64^2$ using the method
described in the text. We show continuous lines for five $\lambda$ values
(bottom to top in the plot): 0.1, 0.2, 0.5, 0.7 and 1.0. 
The vertical line marks
the position of the infinite volume critical temperature. The wider  line
marks the (universal) infinite volume RG prediction ($b_2=2 \tau^2$).
}
\label{fig:predict}
\end{figure} 

To put this program in practice  
we use the value of $C = \pi e^{-2\gamma_E}/2$ 
that we argued was appropriate for lattice regulation,
and we make the fit to the finite size correlator over the range $r=1,32$ since
both our original and new simulations are on a $64^2$ lattice. 
The resulting predictions are shown in figure (\ref{fig:predict}).
These predictions have the same qualitative form as the 
simulation data shown in figure (\ref{fig:old}). Namely,
we observe the same apparent shift in critical temperature with $\lambda$
or appearance of $\log^2$ above $T_c$ and the same dependence of this shift,
which increases with increasing $\lambda$. 


\begin{figure}
\begin{center}
\leavevmode
\epsfig{file=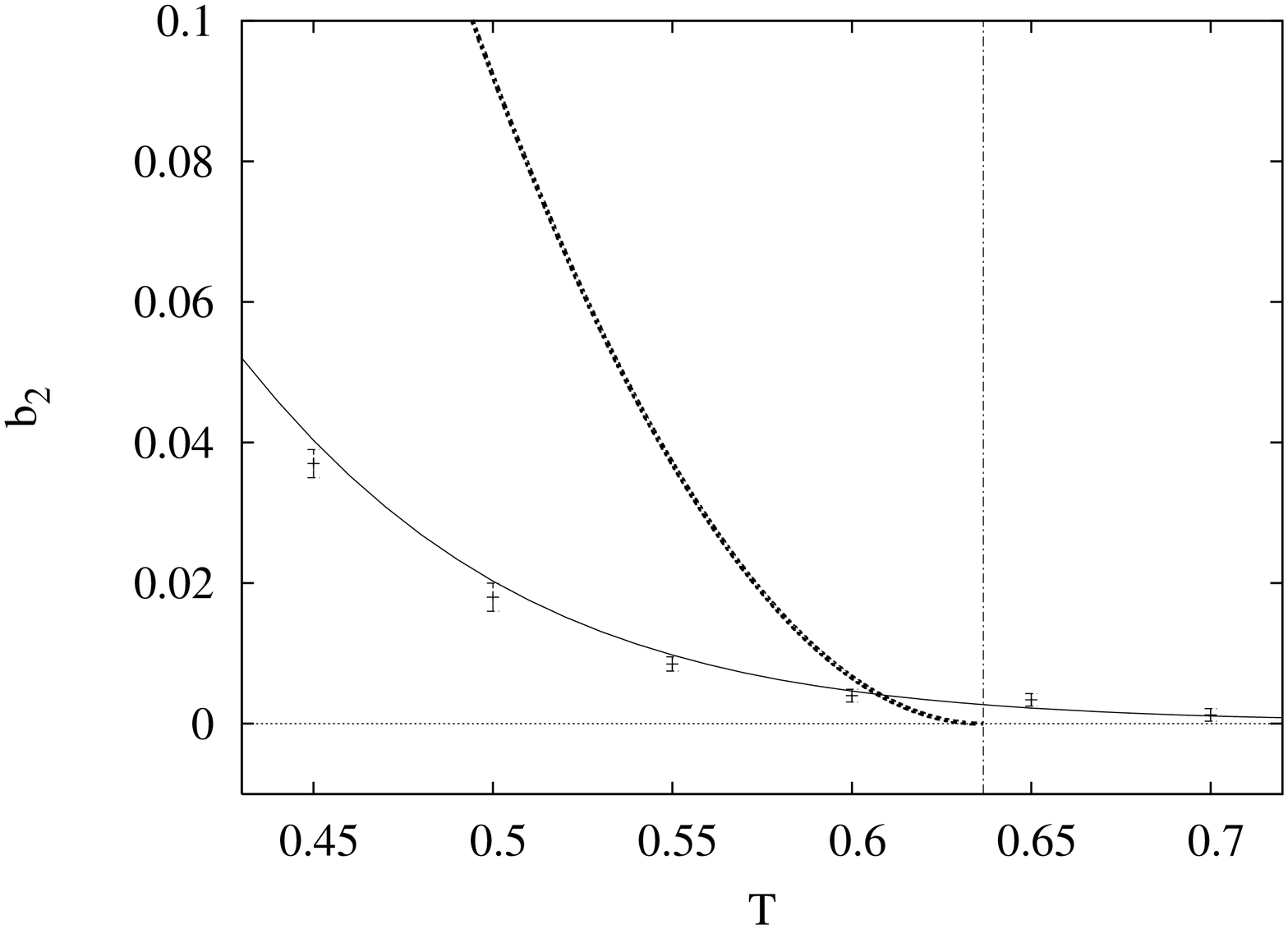,width=0.6\linewidth,angle=270}
\end{center}
\caption{
  Comparison of theoretical estimates for the coefficient $b_2$ of the 
  $\log^2 r$ term in the correlator against new high precision simulations
  at $\lambda=0.5$   in the vicinity of the critical temperature. 
  Axes and marked lines  have the same significance as in earlier figures.
  The continuous line is the theoretical estimate for
  $\lambda=0.5$ and the points are from the new $\lambda=0.5$ numerical
  simulation ($L=64$).}
\label{fig:new}
\end{figure}

For a more precise test of agreement,
figure  (\ref{fig:new}) expands the previous figure 
near the critical temperature to
confront the new high precision $\lambda=0.5$ data against the predictions.
The features identified in the Introduction  
are clearly apparent in the data in figure  (\ref{fig:new}):  
the continued presence of the  $\log^2 r$ 
coefficient above the expected $T_c = 2/\pi = 0.6366$,
and the disagreement with the {\it universal} infinite volume 
prediction for this coefficient below $T_c$.
The similarity between the finite size prediction and data is striking.
Within the temperature range shown, 
the predictions lie within twice the error bound of the simulations.
We can quantify the extent of agreement by
computing  $\chi^2$ between the numerical data and our theoretical
prediction. We obtain $\chi^2=5.13$ using the temperatures in the interval
$[0.5,0.7]$, the number of degrees of freedom (d.o.f.) being 5
(notice that we fit the numerical data to a
function that we have obtained without resort to these data). The probability 
to have a $\chi^2$ larger than $\chi^2=5.13$ for 5 d.o.f. is 40\%, which is
larger than the 5\% limit typically used in the literature to discriminate
the validity of hypotheses \cite{SOKAL}.  If the interval is enlarged 
further below the critical temperature to $[0.45,0.7]$ 
we obtain $\chi^2=7.85$ with 6 d.o.f. corresponding to
a probability of 25\%~\footnote{ We can apply this same procedure to 
  compare the predictions against the original data (dataset A in appendix A). 
  For the 5 d.o.f. interval [0.5,0.7], $\chi^2=14.6$ with 
  $\mathrm{Prob}(\chi^2 > 14.6)=2.4\%$
  and for the 6 d.o.f. interval [0.45,0.7], $\chi^2=13.3$ with 
  $\mathrm{Prob}(\chi^2 >13.3)=2.1\%$. The agreement is 
  considerably less significant than with the new data.}.
This agreement can also be seen by directly comparing the
prediction for the correlator against the result from simulations.
At $T=0.55$ the agreement is very good, with all predicted points lying  
well within the small errors of the data.

In Appendix A, by way of motivating our choice of parameters for
the new simulations, we discuss some exploratory simulations at higher
values of $\lambda$. These suggest that for larger couplings
as we move away from the perturbative regime, the agreement will not be 
so good. Indeed, in figure (\ref{fig:old}) the $\lambda=2.0$
and $\lambda=\infty$ data are not far apart, suggesting that
$\lambda=2.0$ is well into the non-perturbative regime.
For these reasons we did not perform any
further new simulations.
Nonetheless, the predictions in figure (\ref{fig:predict}) 
retain the qualitative
features observed in the original simulations, figure (\ref{fig:old}).


Although our simulations were only at a particular lattice size, it is
illuminating to consider how the results of the procedure above depend on the
lattice size.  We find that above $T_c$, as the size is increased, the
coefficient $b_2$ decreases, but the dependence is extremely weak as was found
in numerical simulations (in references \cite{US1,US2}, $L=64$ and $L=128$
lattices yielded $b_2$ curves with overlapping error bars in the low
temperature phase).  By following our finite size procedure for a range of
lattice sizes up to 512, we find a good fit to a scaling form.  At $T=0.7$ the
coefficient $b_2$ goes to zero according to $1/L^{0.39}$ for both
$\lambda=0.5$ and 1.0.  This exponent takes a value close to $4\tau$ since at
this temperature $\tau=-0.101$.  This result corresponds to a finite size
dependence $b_2(L) \propto g(L)^2$, that might be expected since $g$ enters
the RG equation for $K$ as $g^2$.

\section{Conclusion}

Our main result lies in the similarity shown in figure (\ref{fig:new}) 
between new simulations and the predictions
based on the RG estimate according to our finite size treatment.
Until now, the numerical
simulation results have been a puzzle since it appears that either there is a
$\log^2 r$ component to the correlator above the critical temperature, or that
the location of the critical temperature depends on the disorder.  Neither of
these interpretations are consistent with the conventional RG treatment.  In
this paper we have shown that one can qualitatively understand the numerical
observations in the framework of the RG without the need for any additional
ingredient just by taking account of strong finite size effects.  
Above and near the critical temperature, the disorder
strength renormalises to zero and the fixed point is Gaussian.  However the
dependence on lattice size is extremely slow, and this behaviour induces the
$\log^2 r$ term above the critical temperature.

The statistically significant level of agreement relies on several
factors. Firstly, the use of new  high precision numerical simulations 
that are better thermalized than the old data.
Secondly, the agreement only occurs in the regime where perturbative 
RG is valid, 
namely for small disorder strength ($\lambda \le 0.5$) and near the 
transition $0.5\le T\le 0.7$. 
Finally the level of quantitative agreement is crucially dependent on
the use of  RG coefficients appropriate
to the lattice regulator rather than the traditional Coulomb gas values.

Our treatment does not constitute a full finite size theory,
and there are several factors we have neglected. 
The first of these is the perturbative correction from $g(l)$
to the equation (\ref{Eqcorrg2}).
Although the leading perturbative contributions
to the correlator can be written
down, their numerical evaluation would require additional
approximations that would undermine attempts to make detailed
quantitative comparisons. Moreover, the renormalisation
factor from the anomalous dimension of the vertex operator
has not been computed. 
A proper treatment of these perturbative corrections would
require a full lattice based perturbative computation.
Another factor is the
contribution from irrelevant operators.
Irrelevant operators  can arise from the higher cosine modes after
integrating out the disorder and also from the discretisation.
A brief discussion of these operators is given in appendix C. 
A final factor, the use of an RG coefficient appropriate
to the lattice regulator has been included in our treatment, 
and indeed without this we would not achieve anything like 
the quantitative agreement shown in figure (\ref{fig:new}).

In our earlier numerical work~\cite{US2} we also investigated the dynamics. We
found that the dynamical critical exponent $z$ departs from its Gaussian value
above the critical temperature, and that the value of this term depends on the
disorder strength (see also~\cite{Rieger3}).  We expect that this effect can
also be explained in terms of the same strong finite size effects.

\section*{Acknowledgements}

We wish to thank R. Cuerno, A. S\'anchez and A. Sokal for interesting
discussions.  Partial financial support from CICyT (Spain) grants
BFM2003-08532-C03-02 and FISES2004-01399 is acknowledged. We thank A.
Garc\'{\i}a for providing computing support.


\appendix

\section{Simulation Details}

Here we discuss technical details of both the old and new simulations
and present additional numerical results. All  the data under
discussion has been placed in the Journal repository.

\subsection{Original Simulations (dataset A)}

Our original numerical simulations~\cite{US2} 
were performed in 1995 on an APE supercomputer.
Working on a $64^2$ lattice,
two values of the coupling were considered, $\lambda = 0.5$
and $\lambda = 2.0$ and 128 samples were used for each data point. 
Figure (\ref{fig:old}) also shows
results for the discrete model~\cite{US1,Zeng}, 
but in view of the large coupling this model corresponds to, 
we do not consider it in the present work. 

The data were based on a single annealing run (henceforth called dataset A), 
with $10^5$ thermalization and $2. 10^4$ measurement sweeps at 
each temperature (separated by $\Delta T=0.05$).
The random number generator used in this work~\cite{ParisiRapano}
has since been criticised~\cite{9612024}.


\subsection{New Simulations (datasets B,C)}

To reduce the size of the error bars on dataset A so as to
make quantitative comparisons we performed a number of high precision
simulations around the transition temperature.
Since the APE machine used for the original simulations is no longer
available, we wrote new code running on XEON PC's.
These new simulations remain on the same $64^2$ lattice but
have better thermalization, higher statistics
and an improved random number generator
(a 64-bit congruential generator based on reference~\cite{sokal64bit}). 
Both dataset B and C are at  $\lambda=0.5$.

Dataset B checks the new code by using the same parameters and 
the original annealing schedule of dataset A
and closely reproduces those results.
Error bars generally overlap, though they are slightly apart
at T=0.65,0.7.  A $\chi^2$ evaluation gives a confidence limit of 15\%
that the data arise from the same distribution.
The new data (B) tends to give slightly higher signal than
the old data (A) and we would ascribe this difference, if significant, to the 
improved random number generator. 
We therefore have confidence in the new code.

For improved thermalization, dataset C, the annealing schedule has 
been lengthened and consists of
separate runs of 500 samples with $10^6$ thermalization sweeps at 
temperature intervals of $\Delta T = 0.1$. The final
measurements are then taken over $10^6$ sweeps.
Each data-point (C) lies within two sigma of
the other data, however,
there is a systematic effect that always 
increases the new value of $b_2$ relative to either A or B. 
Thermalization is the main factor in the difference, but we cannot discount an
effect also from the random number generator.

We have also performed runs to observe the approach to
thermalization at $T=0.45$ by tracking the value of the correlator at
maximum distance (32, half the lattice size), as the
measurements progress. This study indicates that 
a thermalization of $10^6$ sweeps is
necessary and sufficient (to an accuracy corresponding to
the number of samples we use).
We are therefore confident in the new data C and this is what we use 
in the main text to 
make quantitative comparisons with the RG predictions.

\subsection{New Exploratory Simulations}

Before deciding the coupling strength to use for run C, we made
a short set of new simulations to explore new values of
$\lambda$ at two fixed temperatures: $T=0.55$ just below
$T_c$ and $T=0.65$ just above $T_c$.

\begin{table*}
\begin{tabular}{|c| c| c|}\hline \hline
$\lambda$ & $b_2^\mathrm{num}$ & $b_2^\mathrm{th}$   \\
\hline
$\infty$ & 0.085(5) & - \\ \hline
2.0    & 0.067(6)      & 0.268  \\ \hline
0.7    & 0.035(2)      & 0.048 \\ \hline
0.6    & 0.031(2)      & 0.033 \\ \hline
0.5    & 0.018(2)      & 0.020 \\ \hline
0.4    & 0.012(1)      & 0.0105 \\ \hline
0.3    & 0.006(1)      & 0.0041 \\ \hline \hline 
\end{tabular}
\caption{Data for $b_2$ just below $T_c$ at $T=0.50$. 
Simulation data (original for $\lambda=\infty,2.0$ other values 
from new simulations) 
denoted by $b_2^\mathrm{num}$ 
compared with predictions 
denoted by $b_2^\mathrm{th}$.}
\label{table_comp0.50}
\end{table*}

Table \ref{table_comp0.50}  shows results
for a variety of $\lambda$ 
just below the critical temperature at $T=0.50$ (128 samples in each case).
At this temperature we are able to make accurate
measurements at small $\lambda$ and have included the data from both new
and old simulations.  We find that for $\lambda\le 0.6$ the predictions 
agree well with the results of the simulations, but at $\lambda=0.7$ 
the value is not statistically compatible.

\begin{table*}
\begin{tabular}{|c| c| c|}\hline \hline
$\lambda$ & $b_2^\mathrm{num}$ & $b_2^\mathrm{th}$   \\
\hline
$\infty$ & 0.025(3)  & -\\ \hline
2.0    & 0.018(4)    & 0.108  \\ \hline
1.0    & 0.0050(15)  & 0.020 \\ \hline
0.7    & 0.0046(17)  & 0.007 \\ \hline
0.5    & 0.0034(9)   &0.002 \\ \hline \hline
\end{tabular}
\caption{Data for $b_2$ just above $T_c$ at $T=0.65$. 
Simulation data (original for $\lambda=\infty,2.0$ other values
from new simulations)  
denoted by $b_2^\mathrm{num}$,
compared with predictions 
denoted by $b_2^\mathrm{th}$.}
\label{table_comp0.65}
\end{table*}

Table \ref{table_comp0.65} shows similar data, based on 500 samples, for a
variety of $\lambda$ just above $T_c$ at $T=0.65$.  For $\lambda \ge 0.5$ the
system shows a clear signal of $\log^2 r$, but we have been unable to obtain
statistically useful results for smaller $\lambda$ due to the number of
samples required.  We find agreement with predictions at $\lambda=0.5$, 
though not at $\lambda=0.7$, matching the range of agreement just above $T_c$.

On the basis of the fixed temperature exploratory simulations it appears that
$\lambda = 0.5$ is within the range of small coupling where the
RG computations are accurate. We therefore 
collected dataset C at fixed  $\lambda=0.5$ over the temperature
range $[0.4,0.7]$.

\section{RG Coefficients by Momentum Shell Integration}

In this appendix we use 
momentum shell integration technique to derive expressions for
the RG coefficients 
that can be used with a
lattice regulator. Although we end up with formulae equivalent
to those obtained using other methods in~\cite{CLD}, it is
only by going through the full derivation that we are able to
check that the expressions remain valid with a lattice regulator.
The momentum shell technique is direct and intuitive
and is an interesting calculation in its own right.
The technique is explained by Kogut for the quite similar (ordered) 
sine-Gordon model in 
\cite{Kogut}. Since we follow this reference quite closely,
in this appendix we only give an outline of the computation with 
particular details of differences from the ordered sine-Gordon case.

\subsection{Notation}

It is convenient to introduce some notation for the local operator:
\begin{equation}
O_{\alpha\beta}(x) 
= e^{2\pi i (\phi_\alpha(x) - \phi_\beta(x))}\,.
\end{equation}
Also for quantities that appear after Gaussian integration.
\begin{eqnarray}
A_{\alpha\beta}(x) 
&=& e^{-2\pi^2 G_h(x) K^{-1}_{\alpha\beta}}\,,\\
B_{\alpha\beta}(x) 
&=& e^{-2\pi^2 G(x) K^{-1}_{\alpha\beta}}\,.
\label{eq:notation}
\end{eqnarray}
Where $G(x)$ is the full propagator of the $\phi$ fields and
$G_h(x)$ is the propagator for the $h$ fields that have
restricted support only in the momentum shell ($\Lambda-d\Lambda<p<\Lambda$).
\begin{equation}
G_h(x) = d\Lambda {dG \over d\Lambda}\,.
\end{equation}
The replica indices only appear through the inverse kinetic
matrix and only take distinct values on or off the diagonal.
It is therefore useful to write: 
\begin{eqnarray}
A(x) &=& A_{\alpha\beta}(x) \quad \alpha=\beta \ 
(\mathrm{diagonal})\,,\\
{\tilde A}(x) &=& A_{\alpha\beta}(x) \quad \alpha\ne\beta \ 
(\mathrm{off-diagonal}) \,.
\end{eqnarray}
And similar expressions for $B$ and $\tilde B$.

Then, using the explicit expression for $K^{-1}_{\alpha\beta}$,
we find a simple form for the combination:
\begin{equation}
A(x) {\tilde A}^{-1}(x) = e^{-2\pi^2 G_h(x)/{\tilde K}}\,.
\end{equation}
In which $K$ does not appear. There is an analogous
expression for the $B$'s.

We also define coefficients 
$b_{\alpha\beta\gamma\delta}(\xi)$ to take account of the
connected expectations at second order, given by:
\begin{equation}
b_{\alpha\beta\gamma\delta}(x)
=
A^2_{\alpha\gamma}(x)A^2_{\beta\delta}(x)A^{-2}_{\alpha\delta}(x)A^{-2}_{\beta\gamma}(x)
-1\,.
\end{equation}

\subsection{Expectations}

The change in action as fields (denoted as $h$) 
in the momentum shell are integrated
out is evaluated perturbatively in $g$.
At first order the Gaussian integration yields the intuitive
contractions:
\begin{eqnarray}
\langle \cos 2\pi(\phi_\alpha(x) - \phi_\beta(x) + h_\alpha(x) - h_\beta(x)) \rangle
&=&
{1\over 2} \left( 
O_{\alpha\beta}(x)\langle e^{2\pi i (h_\alpha(x) - h_\beta(x))}\rangle +\mathrm{c.c.}\right)
\nonumber\\
&=&
A^2(0){\tilde A}^2(0) \cos 2\pi(\phi_\alpha(x) - \phi_\beta(x))\,.
\label{eq:first1}
\end{eqnarray}

The second order term is not much harder. In this and all subsequent
computations we do not worry about operator ordering or normal
ordering and always follow the consistent prescription of the path integral.
\begin{eqnarray}
\langle \cos 2\pi\left(\phi_\alpha(x) - \phi_\beta(x) + h_\alpha(x) - h_\beta(x)\right)&& 
\cos 2\pi\left(\phi_\gamma(y) - \phi_\delta(y) + h_\gamma(y) - h_\delta(y)\right)
\rangle_{\mathrm{connected}} \nonumber\\
&=&
{1\over 2} 
A^4(0){\tilde A}^{-4}(0)
\left( 
b_{\alpha\beta\gamma\delta}(\xi)
O_{\alpha\beta}(x)O_{\gamma\delta}(y)
+\mathrm{c.c.}\right)\,.
\label{eq:second1}
\end{eqnarray}
The cosine operators are at different spatial points $x$ and
$y$ and we have written the difference:  $\xi = x-y$.

The coefficients $b_{\alpha\beta\gamma\delta}(\xi)$ take account of the
connected form of the expectation and were defined above.
For a smooth cutoff, we expect the $b(\xi)$'s to be short 
range and allow us to use an operator product expansion (OPE) for 
$O_{\alpha\beta}(x)O_{\gamma\delta}(y)$.
Indeed, the momentum shell approach requires a regulator function that allows
this step to proceed.
The OPE was not needed in Kogut's work, but as pointed out in
\cite{KnopsDenOuten}, is necessary in general.

\subsection{Operator Product Expansion}

The relevant terms in the OPE are: 
\begin{equation}
O_{\alpha\beta}(x)O_{\gamma\delta}(y)
\sim
a_1(\xi) \delta_{\alpha\delta}\delta_{\beta\gamma}
\left(\partial \phi_\alpha - \partial \phi_\beta\right)^2 
+
a_2(\xi) \left[ 
\delta_{\alpha\delta}(1-\delta_{\beta\gamma})O_{\beta\gamma}
+
\delta_{\beta\gamma}(1-\delta_{\alpha\delta})O_{\alpha\delta}
\right] \,.
\label{eq:ope}
\end{equation}

We compute the coefficients $a_1$ and $a_2$ consistently by
using exactly the same Gaussian integration techniques we
have employed throughout the computation.
\begin{eqnarray}
a_1(\xi) 
&=&
-\pi^2 B^4(0) {\tilde B}^{-4}(0) \xi^2 
B^{-4}(\xi) {\tilde B}^{4}(\xi)\,,\\ 
a_2(\xi) 
&=&
B^2(0) {\tilde B}^{-2}(0)
B^{-2}(\xi) {\tilde B}^{2}(\xi)\,.
\label{eq:opecoeff}
\end{eqnarray}

\subsection{RG coefficients}

After rescaling the action, we identify the changes to
couplings necessary to preserve the physics, and obtain
expressions for the RG coefficients.
\begin{eqnarray}
C &=& 
-(n-2)
\int d^2\xi \, 
a_2(\xi) 
\left[A^{-2}(\xi){\tilde A}^{2}(\xi) - 1 \right]
=
{8\pi^2\over {\tilde K}}
\Lambda^2\int d^2\xi \, 
\Lambda {dG(\xi)\over d\Lambda}
e^{{4\pi^2\over\tilde K} \left[ G(\xi) - G(0)\right]}\,,\\
A &=&
{-2}{\Lambda^4} 
\int d^2\xi \, 
a_1(\xi) 
\left[A^{-4}(\xi){\tilde A}^{4}(\xi) - 1 \right]
=
{16\pi^2\over {\tilde K}}
\Lambda^4\int d^2\xi \, 
\Lambda {dG(\xi)\over d\Lambda}
\xi^2
e^{{8\pi^2\over\tilde K} \left[ G(\xi) - G(0)\right]}\,.
\end{eqnarray}

Notice that these expressions are 
in terms of a regulated propagator and
that although the raw propagator
$G(x)$ is infra-red divergent,
the formulae appear in terms of
well defined propagators: ${dG(\xi)\over d\Lambda}$ and
$G(\xi) - G(0)$.

Having checked that ${dG(\xi)\over d\Lambda}$ is short range
for the lattice propagator, we can be confident in using
the approach with a lattice regulator.

\subsection{Smooth cutoff and Universality}

Recognising that
$G(r)-G(0)$ is a function of the combination $r\Lambda$ 
and that $dG(0)/d\Lambda = 1/2\pi\Lambda$;
we can write the $dG(x)/d\Lambda$ part of the
RG coefficients as:
\begin{equation}
{dG(x)\over d\Lambda} =  
{d\over d\Lambda}\left[ G(x) - G(0)\right] + 
{1\over 2\pi\Lambda}  \,.
\end{equation}
Then re-expressing the $d/d\Lambda$ as a ${\mathitbf{x}.\nabla}$,
the derivative can then be taken to act on the exponential
to obtain (in the spherically symmetric case):
\begin{eqnarray}
C &=& 
{8\pi^2\over {\tilde K}}
\Lambda^2\int d\xi \xi \, 
\left[1 + {{\tilde K}\over 2\pi}\xi{d\over d\xi}\right]
e^{{4\pi^2\over\tilde K} \left[ G(\xi) - G(0)\right]}\,,\\
A &=&
{16\pi^4\over {\tilde K}}
\Lambda^4\int d\xi \xi^3 \, 
\left[1 + {{\tilde K}\over 4\pi}\xi{d\over d\xi}\right]
e^{{8\pi^2\over\tilde K} \left[ G(\xi) - G(0)\right]}\,.
\end{eqnarray}
Now, without taking limits of any kind, write the
propagator in the exponent in the following form for
any value of $r$.
\begin{equation}
G(r) - G(0)
= - {1\over 2\pi} \log (r\Lambda) + {1\over 4\pi} \Phi(r\Lambda)\,. 
\end{equation}
This is a definition of $\Phi$ which contains all the regulator
dependence. For small $r$, this will appear to be an
unhelpful decomposition, since $\Phi(r)$ must contain terms to 
cancel the singularity in the logarithm. 
For large $r$, the logarithm is the leading term
and all remaining regulator dependence comes from $\Phi(\infty)$.
This $\Phi(\infty)$ is the next to leading term of the
asymptotic expansion of the propagator that is merely a 
constant and is often incorporated into the logarithm. 

Evaluating $A$ and $C$ at the critical point (${\tilde K}_c = \pi$):
\begin{eqnarray}
C &=& 
4\pi \int_0^\infty dx {d\over dx} e^{\Phi(x)}
= 
4\pi e^{\Phi(\infty)}\,,\\
A &=&
4\pi^3 \int_0^\infty dx {d\over dx} e^{2\Phi(x)}
= 
4\pi^3 e^{2\Phi(\infty)}\,.
\end{eqnarray}

Our expressions for the RG coefficients confirm the
universal nature of the ratio, since the regulator
dependent pieces contained in $\Phi(\infty)$ cancel
in the ratio. Moreover the ratio takes the correct~\cite{CLD}
value, $R=\pi$.

\section{Irrelevant Operators}

In the theoretical study of the disordered sine Gordon model one
usually picks up only the relevant operators (in the RG
sense). However the irrelevant ones can induce scaling corrections 
which modify the behaviour of the system.
The goal of this section is to identify such
operators and to show up their influence in the scaling of the
numerical observables.

We can identify two main sources of irrelevant operators. The first one
is the discrete nature of the model which leads to:
\begin{equation}
\phi(r+a^\mu)=\phi(r)+ a \partial_\mu \phi+ \frac{1}{2} a^2
\partial_\mu \phi +\frac{1}{6} a^3 \partial^3_\mu \phi + O(a^4)\,,
\end{equation}
where $a^\mu$ is the lattice vector with modulus $a$. 
Hence, in addition to the original kinetic term in the Hamiltonian
$(\partial_\mu \phi)^2)$, the following
terms appear,
\begin{equation}
g_2 \int d^d x\; (\partial_\mu^2 \phi)^2+
g_3 \int d^d x\;(\partial_\mu^3 \phi)^2+
g_4 \int d^d x\; (\partial_\mu \phi) (\partial_\mu^2 \phi)+
g_5 \int d^d x\; (\partial_\mu \phi) (\partial_\mu^3 \phi)+
g_6 \int d^d x\; (\partial_\mu^2 \phi) (\partial_\mu^3 \phi)+O(a^8)\,.
\end{equation}
One can compute the RG equations to the leading order for each coupling 
by simply
computing their (momentum) dimensions using the Gaussian Hamiltonian:
\begin{equation}
\frac{d g_i}{d l}= \left(\mathrm{dim\,} g_i\right) g_i + ...\;,
\end{equation}
with  $\mathrm{dim\,} g_2=-d$, $\mathrm{dim\,}
  g_2=-d$, $\mathrm{dim\, g_3}=-d-2$, $\mathrm{dim\,} g_4=-d+1$,
$\mathrm{dim\,} g_5=-d$, $\mathrm{dim\,} g_6=-d-1$.
Their value at the lattice scale $L$ is given by:
\begin{equation}
g_i(L) \sim L^{-\mathrm{dim\,} g_i}\,.
\end{equation}

The second source of irrelevant operators is the 
disorder average. In this paper we consider a flat distribution 
for the disorder.
In addition to
the two replica term, the full disorder average of the model induces
couplings with an arbitrary number of replicas. The next to leading
terms, with  four replica interactions, are:
\begin{equation}
u_1 \sum_{a b c d} \int d^d x\; \cos 2 \pi(\phi_a+\phi_b-\phi_c-\phi_d)
+u_2 \sum_{a b c d} \int d^d x\; \cos 2 \pi (\phi_a-\phi_b) \cos 2 \pi
(\phi_c-\phi_d) \,.
\end{equation}
By computing their dimensions with the Gaussian Hamiltonian, one can
write the leading order RG equations for $u_1$ and $u_2$:
\begin{equation}
\frac{d u_i}{d l}= 2(2\tau - 1) u_i + ... \,.
\end{equation}
Therefore both of them are irrelevant (from) well below the critical
temperature. At the critical temperature one should expect scaling
corrections proportional to 
$$
u_i(L) \simeq L^{-2}\,.
$$
A similar computation shows that the six replica terms would induce
corrections proportional to $L^{-4}$. In general, an $n$-replica
interaction induces corrections proportional to $L^{-(n-2)}$.
Clearly, all these scaling corrections, are much weaker than
the leading (two replica) one, $g(L)\simeq L^{-2 \tau}$.


\end{document}